\documentclass[conference]{IEEEtran}

\usepackage[utf8]{inputenc}
\usepackage[T1]{fontenc}

\usepackage{amsmath}
\usepackage{amssymb}
\usepackage{booktabs}
\usepackage{dsfont}
\usepackage{enumitem}
\usepackage{pifont}
\usepackage{tabularx}
\usepackage{tikz}
\usetikzlibrary{calc}

\newcommand{\B}{\ensuremath{\mathds{B}}}
\newcommand{\rec}{\ensuremath{\operatorname{rec}}}

\newtheorem{example}{Example}
\newtheorem{theorem}{Theorem}

\usepackage{microtype}

\title{Design Automation and Design Space Exploration for Quantum Computers}
\author{%
  \IEEEauthorblockN{Mathias Soeken$^1$ \qquad Martin Roetteler$^2$ \qquad Nathan Wiebe$^2$ \qquad Giovanni De Micheli$^1$}
  \IEEEauthorblockA{%
    $^1$Integrated Systems Laboratory, EPFL, Lausanne, Switzerland \\
    $^2$Microsoft Research, Redmond, WA, USA
  }
}

\begin{document}

\maketitle

\begin{abstract}
  A major hurdle to the deployment of quantum linear systems algorithms and
  recent quantum simulation algorithms lies in the difficulty to find
  inexpensive reversible circuits for arithmetic using existing hand coded
  methods.  Motivated by recent advances in reversible logic synthesis, we
  synthesize arithmetic circuits using classical design automation flows and
  tools.  The combination of classical and reversible logic synthesis enables
  the automatic design of large components in reversible logic starting from
  well-known hardware description languages such as Verilog. As a prototype
  example for our approach we automatically generate high quality networks for
  the reciprocal $1/x$, which is necessary for quantum linear systems
  algorithms.
\end{abstract}

\section{Introduction}
Quantum computing is getting real.  This year, researchers have fabricated
quantum computers that implement well-known quantum algorithms
reliably~\cite{DLF+16} or perform practical applications such as high-energy
physics simulation~\cite{MMS+16} and electronic structure
computation~\cite{MBK+16}.  Since all such examples involve circuits of very
limited depth, hand designed circuits suffice.  However, as quantum computers
scale up, design automation is necessary in order to fully leverage the power of
this emerging computational model.

Fundamental differences between quantum and classical computing pose serious
design challenges. One is that the basic fault-tolerant gate sets do not include
a universal set of classical gates as fundamental instructions. Instead, one can
implement a universal set of reversible gates by applying a so-called $T$ gate
to the underlying quantum bits (or qubits, or lines). This gate is sufficiently
expensive~\cite{AMMR13} that it is customary to neglect all other gates when
costing a quantum algorithm.  Decomposing the reversible logic that arises in
such algorithms into networks that minimize $T$ gates and qubits is therefore a
central challenge in quantum computing.

Many synthesis algorithms for reversible circuits have been presented in the
last 15 years, see, e.g., \cite{MMD03,MDM07}.  Most of them are applicable to
small functions since they require an explicit function representation, e.g., a
truth table, as input.  In the last few years, more scalable algorithms have
been presented~\cite{SDM16,SC16,PRS15} that work on a symbolic function
representation, thereby allowing reversible circuits to be found for large
functions.

In this paper, we show that scalable reversible logic synthesis algorithms
combined with conventional logic synthesis algorithms allow reversible circuits
to be found automatically for large functions.  We propose design flows that
start from an irreversible design description in Verilog and then use logic
synthesis algorithms to translate it into descriptions that are compatible for
reversible logic synthesis algorithms and finally compile it into a quantum
circuit.  The various algorithms used both in classical and reversible logic
synthesis enable nontrivial design space exploration.  The designer can optimize
the synthesis output with respect to several objectives such as space (number of
qubits), time (number of quantum operations), or runtime of the design flow.  As
a result, the proposed design flows may be robust to changes in quantum
architectures. The design flows further allow researchers to accommodate the
cost of arithmetic and other functions when developing quantum algorithms and
architectures.  To our knowledge, so far such advanced design flows were not
investigated and leveraged for the design of quantum computers.

We illustrate the power of these design flows by finding high quality reversible
implementations of the reciprocal $1/x$ with different bitwidths for $x$.  The
reciprocal is used in several quantum algorithms of high interest.  Most
notably, it is essential for quantum linear systems
algorithms~\cite{HHL09,WBL12}. Recent work has shown that the space requirements
imposed by having to implement the reciprocal reversibly can be prohibitive for
implementations on a small quantum computer~\cite{WR16,BHP+15}.  The
implementation of the reciprocal is used as an example to illustrate the
proposed design flows.  These are the central contributions of this paper and
applicable to many other functions in a similar manner. The uniting advantage is
that a conventional description language such as Verilog can be used as a
starting point.  This enables designers to easily adapt to quantum computing as
well as to easily incorporate a large existing body of conventional logic
synthesis software.

The experimental results confirm the effectiveness of the proposed design flows.
Specifically, we show that we can explore tradeoffs between the number of lines
and the depth of the circuit that cannot be probed using the handcrafted approaches
used in current quantum algorithm design.  This flexibility opens up the possibility of
highly optimized circuits to be introduced to quantum compilation, which allows algorithms
to be better tailored to the severe architectural restrictions imposed by quantum hardware.

\section{Preliminaries}

\subsection{Boolean Functions and Logic Representations}
A multi-output Boolean function $f : \B^n \to \B^m$ maps $n$ Boolean input
values to $m$ Boolean output values and we can represent $f$ as an $m$-tuple of
$n$-variable Boolean functions $(f_1, \dots, f_m)$.  A literal is a Boolean
variable in regular or complemented form.  In logic synthesis, there are several
representations for Boolean functions.  \emph{2-level representations} have a
logic depth of 2; examples are sum-of-products (SOP) in which literals are
combined to product terms using AND and product terms are combined using OR\@.  In
exclusive-sum-of-products (ESOP), the XOR operation is used instead of OR.
2-level representations can be come very large.  \emph{Multi-level
  representations} are directed acyclic graphs called logic networks, in which
terminal nodes are input variables or constants and internal nodes are logic
operations.  In the scope of this paper, we use And-inverter graphs (AIGs,
\cite{Hellerman63}) and XOR-majority graphs (XMGs, \cite{HSA+16}) as logic
networks.  AIGs have AND gates and inverters as logic primitives and XMGs have
XOR, majority-of-three, and inverters as logic primitives~\cite{HSA+16}.

\subsection{Embedding}
Quantum computing requires all operations to be reversible.  This also applies
to the classical Boolean parts, e.g., arithmetic components.  However, many functions of practical interest
are not reversible.  Embedding describes the process of extending an
irreversible $n$-input, $m$-output function $f(x_1, \dots, x_n) = (f_1, \dots, f_m)$
into an $r$-variable reversible function $f'(x_1, \dots, x_r) = (f'_1, \dots, f'_r)$
with $r \ge \max\{n,m\}$.  We say that $f'$ \emph{embeds} $f$, if there exists
assignments $a_{n+1}, \dots, a_r$ such that
\begin{equation}
  \label{eq:emb}
  f_{r-m+j}'(x_1, \dots, x_n, a_{n+1}, \dots, a_r) = f_j(x_1, \dots, x_n)
\end{equation}
for all $1 \le j \le m$.  The assignments $a_{n+1}, \dots, a_r$ are called
\emph{constant inputs}.  The functions $f'_{1}, \dots, f'_{r-m}$ are called
\emph{garbage outputs} as they are discarded in~\eqref{eq:emb} and only required
to make $f'$ reversible.

\begin{theorem}[Bennett embedding, \normalfont{\cite{Bennett73}}]
  \label{thm:bennett}
  Let $f$ be a $n$-input, $m$-output function.  Then
  the $(m+n)$-variable reversible function $f'$ with
  \begin{equation}
    \label{eq:benett}
    f'_j(x_1, \dots, x_{n+m}) =
    \begin{cases}
      x_j & \text{if $j \le n$,} \\
      x_j \oplus f_{j - n}(x_1, \dots, x_n) & \text{otherwise}
    \end{cases}
  \end{equation}
  embeds $f$ for $x_{n+1} \leftarrow 0, \dots, x_{n+m}\leftarrow 0$.
\end{theorem}

The Bennett embedding implies an upper bound on the number of additional lines
$r - n$ that are required to find an embedding.  But for some functions a
smaller number can be found.

The minimum number of required additional lines to embed an irreversible
$n$-input, $m$-output function $f$ is
\begin{equation}
  \label{eq:min-embedding}
  \lceil\log_2\max_{y \in \B^m} \#\{x \in \B^n \mid f(x) = y\} \rceil,
\end{equation}
i.e., the binary logarithm of the maximum size of a collision set of $f$.  An
embedding that ensures this minimum number of additional lines is called
\emph{optimum}.  It has been shown that computing the minimum number of
additional lines is coNP-complete~\cite{SWK+16} and therefore one cannot expect
to find optimum reversible embeddings for large irreversible functions.

\subsection{Reversible Circuits}
In some respects, the structure of a reversible circuit is simpler than the one
of a classical logic network.  A reversible circuit for a function $f : \B^r \to
\B^r$ has $r$ circuit lines on which reversible gates operate which are aligned
in a cascade.  We consider the widely used mixed-polarity multiple-controlled
Toffoli gate (for short Toffoli gate) library in this paper.  Each gate has a
set of control lines that can be positive or negative and one target line that
is disjoint from the control lines.  The gate inverts the value assigned at the
target line if and only if all values assigned to positive (negative) control
lines are 1 (0).  All values on other lines remain unchanged.

\subsection{Reversible Logic Synthesis}
Reversible logic synthesis algorithms can be categorized into \emph{functional
  algorithms} and \emph{structural algorithms}.  \emph{Functional algorithms}
require as input a reversible function and therefore embedding is required as a
preprocessing step prior to synthesis.  Functional algorithms do not add
additional lines to the reversible circuit during synthesis and therefore can
return line optimum results.  Many functional algorithms are based on the
transformation-based approach~\cite{MMD03}, in which Toffoli gates are found
that transform the input function into the identity function, thereby finding a
reversible circuit that realizes the input function.  The original
implementation works on truth tables, but a symbolic variant of the algorithm
exists~\cite{SDM16} that can be applied to larger functions.  For small
functions, SAT-based~\cite{GWDD09} and enumerative approaches~\cite{GFM10} can
even guarantee gate optimum results.

In \emph{structural algorithms} the input function is given in terms of some
structural representation, e.g., a 2-level or multi-level logic network or a
decision diagram.  Synthesis is performed by generating reversible subcircuits
for substructures (e.g., using functional synthesis approaches~\cite{SC16}) and
then concatenating the subcircuits.  Structural algorithms are significantly
more scalable compared to functional ones, but have the drawback of generating a
large number of additional lines---often much higher than the $(m+n)$ bound in
Theorem~\ref{thm:bennett}.

\section{Reciprocal}
We are interested in finding a reversible embedding for the reciprocal function
$ \rec : \B^n \to \B^n$ with $\rec(x_1, \dots, x_n) = (y_1, \dots, y_n)$ such
that $ \frac{1}{x} = y $ when $x = (x_1\dots x_n)_2$ and $y = (0.y_1 \dots
y_n)_2$.  One can increase $n$ to obtain a higher precision with the cost of a
more costly implementation.

We propose two designs to implement the reciprocal using Verilog: (i)
\texttt{INTDIV}$(n)$, which uses Verilog's integer devision operator, and (ii)
\texttt{NEWTON}$(n)$, which implements division using the Newton-Raphson method
on fixed-point numbers.  Both implementations are described in more detail in
the remainder of this section.

\subsubsection{Integer division} We compute the result of the integer division
$2^n / x$, where both $2^n$ and $x$ are represented using $(n+1)$-bit unsigned
integers.  The result is a $(n+1)$-bit unsigned integer from which we omit the
most significant bit.  Then, the remaining bits represent $y$.

\begin{example}
  Let $n = 8$ and let $x = 22$. We have $\frac{1}{22} = 0.0\overline{45}$.  With
  Verilog's integer division, we get $(1\,0000\,0000)_2 / (0\,0001\,0110)_2 =
  (0\,0000\,1011)_2$.  Hence, $y = 2^{-5}+2^{-7}+2^{-8} = 0.04296875$.
\end{example}

\subsubsection{Newton-Raphson method}
We implemented the Newton-Raphson method in Verilog based on signed fixed-point
numbers.  In the following, we use the format Q3.$w$ to denote a signed
fixed-point number in two's-complement encoding that has 3 integer bits
(including the sign bit) and $w$ fractional bits.  Arithmetic operations can be
implemented on signed integer numbers using integer operations.  The result of
an addition or subtraction of two Q3.$w$ numbers is again a Q3.$w$ number.

Multiplication is slightly more involved.  Given a Q3.$w_1$ number $u$ and a
Q3.$w_2$ number $v$, the result of integer multiplication $u * v$ is a Q6.$(w_1
+ w_2)$ number.  We introduce the shortcut $u *_w v$ that truncates the 3 most
significant integer bits and the least significant fractional bits to return a
Q3.$w$ number.

The overall procedure is as follows:
\begin{enumerate}
\item set $x' \leftarrow \text{Q3.$n$}(x / 2^e)$ such that $\frac{1}{2} \le x' <
  1$
\item set $x_0 \leftarrow \text{Q3.$2n$}(48/17) - (\text{Q3.$n$}(32/17) *_{2n}
  x')$
\item for $1 \le i \le I$, \\ \hspace*{10pt}set $x_i \leftarrow x_{i-1} + x_{i-1} *_{2n} \text{Q3.$2n$}(1) - (x' *_{2n} x_{i-1})$
\item set $y' \leftarrow x_I \gg e$
\item set $y$ to the $n$ most significant bits of $y'$
\end{enumerate}

In step 1, we normalize $x$ and make it a fixed-point number $x'$ such that the
integer part is $0$ and the most significant fractional bit is 1.  Note that
this can be done using a right-shift by $e$.  In step 2, we compute the starting
value $x_0$ using constants $48/17$ and $32/17$.  We apply the Newton iteration
$I = \lceil \log_2\frac{P+1}{\log_2 17}\rceil$ times.  The values $x_i$ for $0
\le i \le I$ have $2n$ fractional bits, i.e., we use twice the input precision
to carry out the computations in the Newton iteration (step 3).  We use the same
exponent $e$ to shift the value of $x_I$ and then extract the $n$ most
significant bits as $y$.

\section{Design Flows}

\begin{figure}[t]
  \centering
  \includegraphics{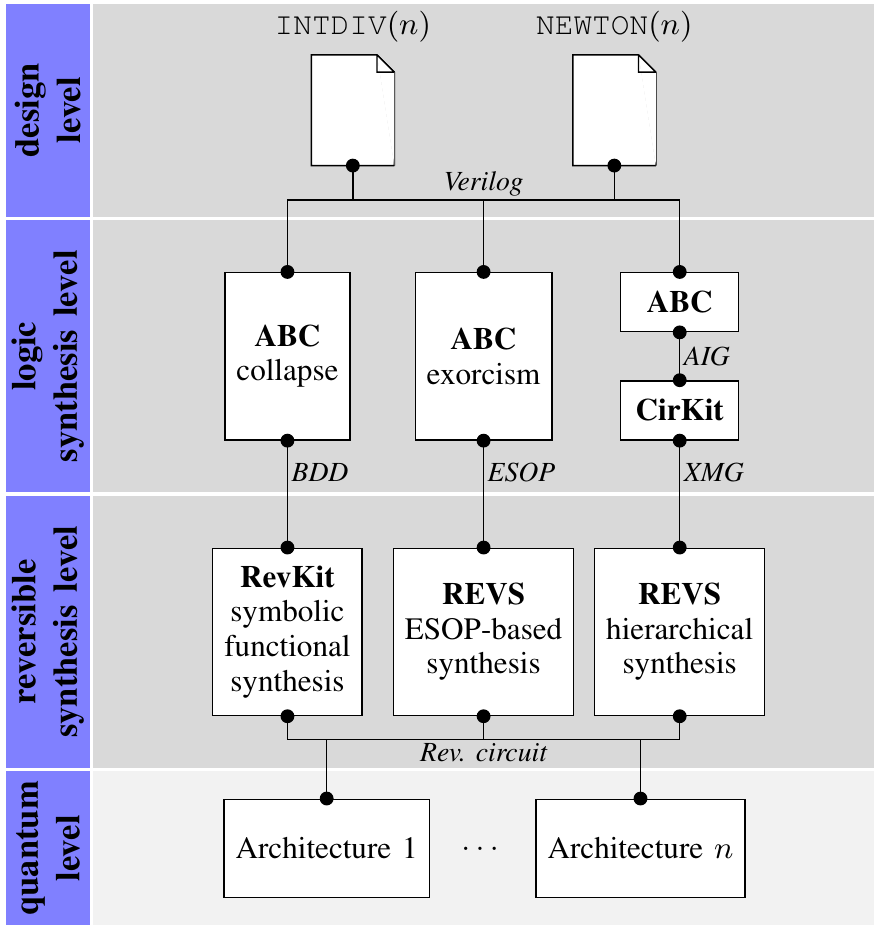}
  \caption{Design flows; lines represent interfaces between files and tools and
    are annotated using the function representation at the interface}
  \label{fig:design-flows}
\end{figure}

This section describes the main contribution of the paper.  We show a variety of
design flows starting from the two Ve\-ri\-log designs \texttt{INTDIV}$(n)$ and
\texttt{NEWTON}$(n)$ that have been introduced in the previous section.
Fig.~\ref{fig:design-flows} offers an overview of the design flows which pass
four levels: (i) the \textit{design level} containing the Verilog descriptions,
(ii) the \textit{logic synthesis level} in which the designs are optimized and
transformed into formats required by (iii) the \textit{reversible synthesis
  level} in which synthesis algorithms generate reversible networks that can
eventually be mapped to (iv) architectures at the \textit{quantum level}.  In
the scope of this paper, we stop after the reversible networks have been
obtained.  The experiments in the next section will show that \texttt{INTDIV}$(n)$ is
superior to \texttt{NEWTON}$(n)$ in this experiment, both in quality and runtime.
However, a simple design such as \texttt{INTDIV}$(n)$ is not possible for some
functions.  Functions such as $\frac{1}{\sqrt{x}}$ or trigonometric functions
require approximation techniques with an implementation similar to the
\texttt{NEWTON}$(n)$ design.  Thus \texttt{NEWTON}$(n)$ can be considered a proxy  for reversible synthesis of other functions.

At the reversible synthesis level we consider synthesis algorithms from three
different approach categories to target different cost aspects: (i)
\textit{symbolic functional synthesis} for a low number of qubits, (ii)
\textit{ESOP-based synthesis} for moderate number of qubits, and (iii)
\textit{hierarchical synthesis} as a scalable solution for large bitwidths and
low $T$-count.  How each algorithm is used in the flow is described in more
detail in the following.

\subsection{Symbolic Functional Synthesis}
The input to symbolic functional synthesis is a binary decision diagram (BDD).
This is obtained by reading the Verilog into the logic synthesis tool
ABC~\cite{BM10} optimize it several rounds using `\emph{dc2}' before collapsing
it into a BDD using `\emph{collapse}'.

At the reversible synthesis level an optimum embedding is obtained from the BDD
(see~\cite{SWK+16,SDM16}).  The resulting reversible function is represented as
a BDD that is input to the SAT-based variant of the symbolic
transformation-based synthesis algorithm~\cite{SDM16}.  The inputs $x_1, \dots,
x_n$ are not preserved by the embedding, i.e., the function $\rec(x_1, \dots,
x_n)$ is applied \emph{in-place}.  This synthesis algorithm guarantees the
optimum number of qubits which must be in between $n$ and $2n$.  A property of
the transformation-based algorithm is that large Toffoli gates with controls on
all circuits lines are generated.  This leads to large $T$-count.

\subsection{ESOP-based Synthesis}
The input is a multi-output ESOP expression.  This is obtained with ABC by first
optimizing the input design using the command sequence `\emph{satclp; sop; fx;
  strash; dc2}' which generates an AIG\@.  We then use `\emph{\&exorcism}' to
collapse the AIG into a ESOP and optimize it using the algorithm presented
in~\cite{MP01}. We then perform ESOP-based synthesis with the reversible
  synthesis tool REVS \cite{PRS15} which is a tool that allows to trade off
between circuit size and the number of qubits. REVS offers different strategies
for cleaning up intermediate calculations and re-using the qubits that have been
freed up. For Boolean functions in ESOP format, REVS offers a mode for factoring
common subexpressions, the size of which are bounded by an integer parameter
$p$. If $p=0$, then REVS proceeds by taking each product term with $k$ literals
in the given multi-output ESOP expression and translates it into a Toffoli gate
with $k$ controls and polarities according to the literals' polarities.  If a
product term is shared among several outputs, only one Toffoli gate is required
to realize one output and CNOT gates are used to copy the result to the other
outputs.  Then the synthesis algorithm creates circuits with $2n$ qubits, and
each Toffoli gate cannot have more than $n$ controls. If $p>0$, then REVS
considers groups of $p$ product terms in the ESOP expansion that have the same
output value and tries to factors the resulting expressions. Intermediate
results in the factorization are stored in additional lines, which typically
leads to an increase of the total number of used lines beyond $2n$. However,
overall this often leads to a reduction of the overall number of $T$ gates, in
particular if many terms in the ESOP expansion correspond to the exact same
function value.

\subsection{Hierarchical Synthesis}
We perform hierarchical synthesis with REVS by using as input an
\emph{XOR-Majority Graph} (XMG).  An XMG is a logic network in which the
primitives are XOR, AND, OR, the MAJ (majority-of-three function, see, e.g.,
\cite{AGM14}), and their inverted forms.  This network representation is
advantageous for reversible logic synthesis.  First, the MAJ gate can be
realized with only one Toffoli gate and therefore has the same number of
$T$ gates as an AND and OR gate by being more expressive.  Second, an XOR gate
can be realized using CNOT gates and therefore does not require any $T$ gates.
Third, the XOR gate can be applied in-place, if the value of at least one its
operands is no longer required to compute another gate.  The same applies for
the MAJ gate, if all of its operands are no longer required.  We derive
optimized XMGs from optimized AIGs using the algorithm presented
in~\cite{HSA+16} using CirKit's\footnote{github.com/msoeken/cirkit} command
`\emph{xmglut -k 4}' on AIGs that were optimized using multiple iterations of
`\emph{resyn2}' in ABC.

\begin{table}[t]
  \caption{Baseline results with manual design}
  \label{tbl:qnewton}
  \catcode`\_=\active\def_{\hskip1ex}%
  \begin{tabularx}{\linewidth}{Xrrrr}
    \toprule
    & \multicolumn{2}{c}{\texttt{RESDIV$(n)$}} & \multicolumn{2}{c}{\texttt{QNEWTON$(n)$}} \\
    \midrule
    $n$ & qubits & $T$-count & qubits & $T$-count \\[3pt]
    __8 & 48   & 8\,512   & 111  & 14\,632     \\
    _16 & 96   & 34\,944  & 234  & 64\,004     \\
    _32 & 192  & 141\,568 & 615  & 352\,440    \\
    _64 & 384  & 569\,856 & 1226 & 1\,405\,284 \\
    \bottomrule
  \end{tabularx}
\end{table}

\begin{table}[t]
  \def\tabcolsep{4pt}
  \caption{Results with symbolic functional reversible synthesis}
  \label{tbl:pla}
  \catcode`\_=\active\def_{\hskip1ex}%
  \begin{tabularx}{\linewidth}{Xrrrrrr}
    \toprule
    & \multicolumn{3}{c}{\texttt{INTDIV$(n)$}} & \multicolumn{3}{c}{\texttt{NEWTON$(n)$}} \\
    \midrule
    $n$ & qubits & $T$-count & runtime & qubits & $T$-count & runtime \\[3pt]
_4 & 7 & 597 & 0.10 & 7 & 589 & 0.29\\
_5 & 9 & 1\,613 & 0.11 & 9 & 1\,848 & 0.32\\
_6 & 11 & 5\,963 & 0.18 & 11 & 6\,419 & 0.41\\
_7 & 13 & 20\,008 & 0.31 & 13 & 17\,867 & 0.57\\
_8 & 15 & 51\,386 & 0.74 & 15 & 56\,379 & 4.15\\
_9 & 17 & 142\,901 & 2.54 & 17 & 148\,913 & 6.47\\
10 & 19 & 380\,009 & 10.94 & 19 & 383\,891 & 15.72\\
11 & 21 & 946\,724 & 43.62 & 21 & 945\,117 & 54.44\\
12 & 23 & 2\,318\,841 & 284.72 & 23 & 2\,346\,319 & 296.67\\
13 & 25 & 5\,599\,538 & 1\,862.22 & 25 & 5\,645\,530 & 1\,669.84\\
14 & 27 & 13\,148\,102 & 10\,545.50 & 27 & 13\,186\,076 & 9\,342.96\\
15 & 29 & 30\,761\,399 & 44\,501.20 & 29 & 30\,746\,528 & 41\,398.30\\
16 & 31 & 71\,155\,258 & 274\,744.00 & 31 & 71\,259\,272 & 213\,135.52\\
    \bottomrule
  \end{tabularx}
\end{table}

\begin{table*}[t]
  \def\tabcolsep{6.5pt}
  \caption{Results with REVS~\cite{PRS15}}
  \label{tbl:esop}
  \catcode`\_=\active\def_{\hskip1ex}%
  \begin{tabularx}{\linewidth}{Xrrrrrrrrrrrr}
    \toprule
    & \multicolumn{3}{c}{\texttt{INTDIV$(n)$, $p=0$}} & \multicolumn{3}{c}{\texttt{NEWTON$(n)$, $p=0$}} & \multicolumn{3}{c}{\texttt{INTDIV$(n)$, $p=1$}} & \multicolumn{3}{c}{\texttt{NEWTON$(n)$, $p=1$}} \\
    \midrule
    $n$ & qubits & $T$-count & runtime & qubits & $T$-count & runtime & qubits & $T$-count & runtime & qubits & $T$-count & runtime \\[3pt]
_5 & 10 & 232 & 0.04 & 10 & 135 & 0.26 & 12 & 241 & 0.04 & 10 & 135 & 0.26\\
_6 & 12 & 423 & 0.09 & 12 & 294 & 0.35 & 14 & 411 & 0.08 & 13 & 268 & 0.32\\
_7 & 14 & 791 & 0.08 & 14 & 568 & 0.35 & 17 & 803 & 0.11 & 17 & 511 & 0.34\\
_8 & 16 & 1\,342 & 0.12 & 16 & 1\,039 & 3.92 & 20 & 1\,349 & 0.13 & 20 & 1\,060 & 3.94\\
_9 & 18 & 2\,056 & 0.19 & 18 & 1\,894 & 5.62 & 23 & 1\,887 & 0.21 & 22 & 1\,850 & 5.68\\
10 & 20 & 3\,415 & 0.32 & 20 & 3\,311 & 9.27 & 23 & 3\,238 & 0.32 & 24 & 3\,071 & 9.16\\
11 & 22 & 5\,631 & 0.52 & 22 & 5\,303 & 14.98 & 29 & 5\,244 & 0.69 & 27 & 4\,846 & 14.97\\
12 & 24 & 8\,431 & 0.95 & 24 & 8\,423 & 25.09 & 30 & 7\,700 & 1.20 & 29 & 7\,136 & 25.29\\
13 & 26 & 13\,414 & 1.93 & 26 & 14\,287 & 44.03 & 31 & 11\,474 & 2.01 & 32 & 11\,988 & 44.16\\
14 & 28 & 21\,902 & 3.22 & 28 & 21\,782 & 72.70 & 37 & 19\,063 & 3.82 & 34 & 19\,186 & 72.94\\
15 & 30 & 33\,502 & 8.18 & 30 & 34\,815 & 118.74 & 35 & 27\,897 & 8.51 & 37 & 28\,635 & 118.99\\
16 & 32 & 52\,376 & 19.13 & 32 & 50\,784 & 1\,128.73 & 46 & 42\,717 & 1\,598.71 & 38 & 41\,532 & 1\,129.66\\
17 & 34 & 78\,470 & 36.24 & 34 & 95\,462 & 1\,860.05 & 41 & 64\,089 & 37.46 & 43 & 76\,022 & 1\,861.80\\
18 & 36 & 119\,510 & 86.64 & 36 & 153\,414 & 3\,182.44 & 43 & 94\,577 & 101.05 & 44 & 119\,657 & 3\,186.19\\
19 & 38 & 179\,095 & 169.99 & 38 & 229\,768 & 5\,276.51 & 46 & 138\,912 & 182.24 & 46 & 175\,598 & 5\,286.83\\
20 & 40 & 284\,118 & 393.96 & 40 & 349\,398 & 11\,486.74 & 48 & 218\,341 & 440.79 & 47 & 263\,106 & 11\,502.31\\
21 & 42 & 422\,806 & 823.56 & 42 & 552\,496 & 18\,869.36 & 51 & 318\,627 & 852.37 & 51 & 412\,488 & 18\,936.66\\
22 & 44 & 640\,351 & 3\,075.21 & 44 & 837\,646 & 29\,371.18 & 52 & 476\,603 & 3\,137.62 & 53 & 616\,065 & 30\,666.43\\
23 & 46 & 941\,408 & 7\,462.87 & 46 & 1\,249\,894 & 52\,547.84 & 56 & 684\,166 & 7\,631.37 & 56 & 909\,364 & 52\,936.22\\
24 & 48 & 1\,417\,327 & 19\,487.67 & 48 & 1\,885\,742 & 106\,612.57 & 57 & 1\,021\,041 & 19\,915.68 & 58 & 1\,344\,400 & 107\,490.99\\
25 & 50 & 2\,119\,663 & 72\,035.25 & 50 & 2\,819\,902 & 220\,349.12 & 60 & 1\,512\,893 & 73\,999.97 & 60 & 1\,985\,367 & 222\,364.17\\
    \bottomrule
  \end{tabularx}
\end{table*}

\section{Experiments}
We have generated reversible circuits for the \texttt{INTDIV}$(n)$ and
\texttt{NEWTON}$(n)$ designs using all three design flows.  This section
presents the results of the experimental evaluation.  We used the command
`\emph{tbs -s}' in RevKit~\cite{SFWD12} for the symbolic functional
synthesis~\cite{SDM16}.  We used REVS~\cite{PRS15} to obtain results for both
the ESOP-based and hierarchical synthesis algorithm. All experiments have been
carried out on an Intel E5-2670 octacore CPU with 2.60 GHz and 128 GB main
memory. All runtimes are given in seconds. Correctness of the synthesized
designs has been verified using ABC's combinational equivalence checker
`\textit{cec}'.  All generated files and pointers to implementation details can
be found at \textit{msoeken.github.io/reciprocal.html}.

We use two manual quantum circuit designs for a baseline comparison. First, an
integer division algorithm based on the restoring division
algorithm~\cite{TVM16} that computes for $n$-bit inputs $a$ and $b$ the $n$-bit
quotient $q$ and $n$-bit remainder $r$ such that $a = qb + r$, using $3n$
qubits.  We refer to this circuit as \texttt{RESDIV}.  One can use the circuit
to compute the $n$-bit reciprocal $1/x$ by setting $a = 2^n$ and $b=x$ in a
$2n$-bit version of the circuit in order to match the precision of our designs.
We also manually created a design following the Newton-Raphson method, which
is similar but more accurate to the designs proposed in~\cite{WR16}
  and~\cite{BHP+15} and we refer to it as \texttt{QNEWTON}.  In contrast to the
\texttt{NEWTON} design that follows the standard algorithm, we adjusted the
algorithm as follows to reduce the number of lines needed.  \texttt{QNEWTON}
works by first bitshifting the inputs into the range $[0.5,1)$, implementing
Newton iterations with the Cucarro adder~\cite{CDK+04}, text book
multiplication, and then finally bit shifting the values again to provide the
desired answer.  The precision of the adders used were varied at each Newton
iteration to minimize the space and time resources needed to hit the target
accuracy.

\texttt{QNEWTON}'s use of variable internal precision for the Newton iterations
allows us to compute with roughly half the qubits predicted in previous results
that examined computing reciprocals on quantum computers using Newton
iterations~\cite{WR16,BHP+15}. Our work also differs from~\cite{WR16,BHP+15} in
that here detailed $T$ gate estimates are provided.  As such, these numbers are
a slight improvement upon the previous state of the art.

For each reversible circuit we report the number of qubits, the $T$-count
(according to~\cite{Maslov16} and~\cite{BBC+95}), and the overall runtime of the
flow.  The baseline results obtained from $\texttt{RESDIV}(n)$ and
$\texttt{QNEWTON}(n)$ are given in Table~\ref{tbl:qnewton} for $n = 8, 16, 32,
64$.

Table~\ref{tbl:pla} lists the experimental results for $n \le 16$ when using
symbolic functional synthesis.  For these circuits the number of qubits is
optimum.  That the numbers are equivalent for \texttt{INTDIV} and
\texttt{NEWTON} is not necessarily expected, as \texttt{NEWTON} approximates
$1/x$ which may have an effect on the maximum occurrence of an output
assignment.  The number of qubits is $3.2\times$ and $3.1\times$ smaller
compared to the \texttt{RESDIV} baseline for $n=8$ and $n=16$, respectively.
However, this comes with the price of a very high $T$-count.  The numbers for
\texttt{NEWTON} are slightly higher compared to \texttt{INTDIV} with exceptions
in case of $n \in \{4, 7, 11, 15\}$.  The reason for this large number is that
functional synthesis generated reversible circuits with Toffoli gates that have
a large number of control lines.  For example, the realizations for $n=16$
contain Toffoli gate with up to 27 control lines.  For \texttt{INTDIV}, the
$T$-count is $6.0\times$ and $2036.3\times$ larger compared to \texttt{RESDIV}
for $n=8$ and $n=16$.  The comparison to \texttt{QNEWTON} is qualitatively
similar. The runtimes are very high reaching about 3.2 days for $n=16$ making
this design flow not scalable for larger bitwidths. Despite the increased
runtimes, this result is remarkable because it shows that our design flow can
find designs that use less than the $2n$ lines required for the out of place
reciprocal circuit.

Table~\ref{tbl:esop} lists results for $n \le 25$ when using REVS with a 2-level
ESOP description as input. For $p=0$, the number of qubits is $2n$ which is only
one qubit more compared to the functional synthesis approach.  Compared to the
baseline the number of qubits is $3\times$ smaller for both $n=8$
and $n=16$.  However, as the Toffoli gates have fewer number of controls, the
$T$-count is much smaller compared to functional synthesis.  For small $n$ the
\texttt{NEWTON} design has better $T$-count, which changes for large $n$.  When
comparing the \texttt{NEWTON} design to the \texttt{RESDIV} baseline, the
$T$-count is $8.2\times$ better for $n=8$ but $1.5\times$ larger for
$n=16$.  Relative to \texttt{QNEWTON}, we see comparable numbers of $T$ gates
and far fewer lines at $p=0$, however the ESOP-based approach in REVS
outperforms it at $p=1$.  The REVS-based design flow is more scalable than the
functional approach, but also reaches its limits: for $n=25$, it takes about 20
hours to find a realization for the \texttt{INTDIV} design, and about 2.5 days
for the \texttt{NEWTON} design.

\begin{table}[t]
  \caption{Results with hierarchical synthesis}
  \label{tbl:dxs}
  \catcode`\_=\active\def_{\hskip1ex}%
  \def\tabcolsep{5pt}
  \begin{tabularx}{\linewidth}{Xrrrrrr}
    \toprule
    & \multicolumn{3}{c}{\texttt{INTDIV$(n)$}} & \multicolumn{3}{c}{\texttt{NEWTON$(n)$}} \\
    \midrule
    $n$ & qubits & $T$-count & runtime & qubits & $T$-count & runtime \\[3pt]
    _16 &      892 &   5\,607 &   1.67 &  10\,713    &  73\,080    &      75.66 \\
    _32 &   3\,501 &  21\,455 &  15.48 &  56\,207    & 392\,917    &  1\,218.28 \\
    _64 &  13\,465 &  80\,339 &  38.34 & 178\,653    & 1\,264\,704 &  3\,008.98 \\
    128 &  51\,897 & 308\,364 & 376.39 & 1\,029\,441 & 7\,033\,040 & 37\,575.67 \\
    \bottomrule
  \end{tabularx}
\end{table}

Table~\ref{tbl:dxs} lists results when using the hierarchical synthesis
approach.  This approach can scale to large bitwidths as can be seen from the
\texttt{INTDIV} design.  We show results up to $n=128$, but circuits for larger
bitwidths can still be obtained in a reasonable amount of time.  First, we like
to point out that the results for \texttt{INTDIV} differ significantly from
\texttt{NEWTON}, in contrast to the other two design flows.  This is due to the
fact, that we perform logic optimization at AIG level after the network has been
synthesized from its Verilog description.  The starting points are significantly
different and optimization approaches can easily get stuck in local minima.
Drastic measures such as collapsing the network into a 2-level logic form (as in
the two previous design flows) are required in order to escape from them.
However, collapsing does not scale to these high bitwidths.  Due to this large
difference in quality, we use \texttt{INTDIV} for comparison to the baseline
design.  The number of qubits is $9.3\times$ and $18.2\times$ larger for $n=16$
and $n=32$ compared to the \texttt{RESDIV} baseline design.  However, the
$T$-count is $6.2\times$ and $6.6\times$ smaller for $n=16$ and $n=32$.  Both,
the number of qubits and the number of $T$-count can be improved by spending
more effort in minimizing the number of gates in the XMGs during logic
synthesis, with the cost of a higher runtime.  The $T$-count of \texttt{NEWTON}
also is comparable to \texttt{QNEWTON} however the latter requires $46\times$
and $91\times$ fewer lines for $n=16$ and $n=32$.  This discrepancy occurs
because the hierarchical approach does not directly optimize the precision in
each Newton iteration. Although \texttt{INTDIV} shows better performance in this
example, the numbers for \texttt{NEWTON} are still quite meaningful.  As
discussed above, for functions such as $\frac{1}{\sqrt{x}}$ or trigonometric
functions Newton's method will frequently be the technique of choice for logic
synthesis.  These designs are therefore meaningful benchmarks for
\texttt{NEWTON}'s performance for other logic synthesis problems.

\section{Conclusions}
We presented versatile design flows for the synthesis of reversible logic in
quantum computers.  Our flows take Verilog programs as input that are translated
using classical logic synthesis algorithms into formats appropriate for
reversible logic synthesis algorithms. This enables design exploration and gives
the designer the possibility to optimize with respect to a cost metric such as
the number of $T$ gates or qubits, metrics that correspond to time and space in
quantum computers.  These capabilities are absent in existing approaches for
quantum circuit compilation.  Our work provides a necessary tool for making
quantum algorithms practical, such as quantum linear systems algorithms and
quantum simulation algorithms.

We illustrated the design flows and synthesize a variety of reversible circuits
for the reciprocal $1/x$ with different bitwidths for $x$ and show that we are
able to find circuits that beat handcrafted designs in either width or size,
depending on our optimization goal. In future work we plan to integrate our
design flows into industrial logic synthesis software and find efficient
reversible implementations for floating point arithmetic designs.

\subsubsection*{Acknowledgments} The authors wish to thank Thomas H\"aner, Alan
Mishchenko, Alex Parent, and the anonymous reviewers for their helpful comments.
This research was supported by the European Research Council (H2020-ERC-2014-ADG
669354 CyberCare) and the Swiss National Science Foundation (200021\_169084
MAJesty).

\bibliographystyle{IEEEtran}
\bibliography{library}

\begin{thebibliography}{10}
\providecommand{\url}[1]{#1}
\csname url@samestyle\endcsname
\providecommand{\newblock}{\relax}
\providecommand{\bibinfo}[2]{#2}
\providecommand{\BIBentrySTDinterwordspacing}{\spaceskip=0pt\relax}
\providecommand{\BIBentryALTinterwordstretchfactor}{4}
\providecommand{\BIBentryALTinterwordspacing}{\spaceskip=\fontdimen2\font plus
\BIBentryALTinterwordstretchfactor\fontdimen3\font minus
  \fontdimen4\font\relax}
\providecommand{\BIBforeignlanguage}[2]{{%
\expandafter\ifx\csname l@#1\endcsname\relax
\typeout{** WARNING: IEEEtran.bst: No hyphenation pattern has been}%
\typeout{** loaded for the language `#1'. Using the pattern for}%
\typeout{** the default language instead.}%
\else
\language=\csname l@#1\endcsname
\fi
#2}}
\providecommand{\BIBdecl}{\relax}
\BIBdecl

\bibitem{DLF+16}
S.~Debnath, N.~M. Linke, C.~Figgatt, K.~A. Landsman, K.~Wright, and C.~Monroe,
  ``Demonstration of a small programmable quantum computer with atomic
  qubits,'' \emph{Nature}, vol. 536, pp. 63--66, 2016.

\bibitem{MMS+16}
E.~A. Martinez, C.~A. Muschik, P.~Schindler, D.~Nigg, A.~Erhard, M.~Heyl,
  P.~Hauke, M.~Dalmonte, T.~Monz, P.~Zoller, and R.~Blatt, ``Real-time dynamics
  of lattice gauge theories with a few-qubit quantum computer,'' \emph{Nature},
  vol. 534, pp. 516--519, 2016.

\bibitem{MBK+16}
P.~J.~J. O'Malley \emph{et~al.}, ``Scalable quantum simulation of molecular
  energies,'' \emph{Phys. Rev. X}, vol.~6, p. 031007, 2016.

\bibitem{AMMR13}
M.~Amy, D.~Maslov, M.~Mosca, and M.~Roetteler, ``A meet-in-the-middle algorithm
  for fast synthesis of depth-optimal quantum circuits,'' \emph{{IEEE} Trans.
  {CAD} of Int. Circ. and Syst.}, vol.~32, no.~6, pp. 818--830, 2013.

\bibitem{MMD03}
D.~M. Miller, D.~Maslov, and G.~W. Dueck, ``A transformation based algorithm
  for reversible logic synthesis,'' in \emph{DAC}, 2003, pp. 318--323.

\bibitem{MDM07}
D.~Maslov, G.~W. Dueck, and D.~M. Miller, ``{Techniques for the synthesis of
  reversible Toffoli networks},'' \emph{{ACM} Trans. Design Autom. Electr.
  Syst.}, vol.~12, no.~4, 2007.

\bibitem{SDM16}
M.~Soeken, G.~W. Dueck, and D.~M. Miller, ``A fast symbolic transformation
  based algorithm for reversible logic synthesis,'' in \emph{Int'l Conf. on
  Reversible Computation}, 2016.

\bibitem{SC16}
M.~Soeken and A.~Chattopadhyay, ``Unlocking efficiency and scalability of
  reversible logic synthesis using conventional logic synthesis,'' in
  \emph{Proc. DAC}, 2016, pp. 149:1--149:6.

\bibitem{PRS15}
A.~Parent, M.~Roetteler, and K.~M. Svore, ``Reversible circuit compilation with
  space constraints,'' \emph{arXiv}, vol. 1510.00377, 2015.

\bibitem{HHL09}
A.~W. Harrow, A.~Hassidim, and S.~Lloyd, ``Quantum algorithm for linear systems
  of equations,'' \emph{Phys.~Rev.~Lett.}, vol. 103, no.~15, p. 150502, 2009.

\bibitem{WBL12}
N.~Wiebe, D.~Braun, and S.~Lloyd, ``Quantum algorithm for data fitting,''
  \emph{Physical review letters}, vol. 109, no.~5, p. 050505, 2012.

\bibitem{WR16}
N.~Wiebe and M.~Roetteler, ``{Quantum arithmetic and numerical analysis using
  Repeat-Until-Success circuits},'' \emph{Quantum Information and
  Communication}, vol.~16, pp. 134--178, 2016.

\bibitem{BHP+15}
M.~K. Bhaskar, S.~Hadfield, A.~Papageorgiou, and I.~Petras, ``Quantum
  algorithms and circuits for scientific computing,'' \emph{arXiv:1511.08253},
  2015.

\bibitem{Hellerman63}
L.~Hellerman, ``A catalog of three-variable {Or}-invert and {And}-invert
  logical circuits,'' \emph{{IEEE} Trans. Electronic Computers}, vol.~12,
  no.~3, pp. 198--223, 1963.

\bibitem{HSA+16}
W.~Haaswijk, M.~Soeken, L.~G. Amarù, P.-E. Gaillardon, and G.~De~Micheli, ``A
  novel basis for logic rewriting,'' in \emph{Asia and South Pacific Design
  Automation Conference}, 2017.

\bibitem{Bennett73}
C.~H. Bennett, ``Logical reversibility of computation,'' \emph{IBM Jrnl. of
  Research and Development}, vol.~17, pp. 525--532, 1973.

\bibitem{SWK+16}
M.~Soeken, R.~Wille, O.~Keszocze, D.~M. Miller, and R.~Drechsler, ``Embedding
  of large {Boolean} functions for reversible logic,'' \emph{J. Emerg. Techn.
  in Comp. Systems}, vol.~12, no.~4, p.~41, 2016.

\bibitem{GWDD09}
D.~Gro{\ss}e, R.~Wille, G.~W. Dueck, and R.~Drechsler, ``Exact multiple-control
  {Toffoli} network synthesis with {SAT} techniques,'' \emph{{IEEE} Trans.
  {CAD} of Int. Circ. and Syst.}, vol.~28, no.~5, pp. 703--715, 2009.

\bibitem{GFM10}
O.~Golubitsky, S.~M. Falconer, and D.~Maslov, ``Synthesis of the optimal 4-bit
  reversible circuits,'' in \emph{Proc. DAC}, 2010, pp. 653--656.

\bibitem{BM10}
R.~K. Brayton and A.~Mishchenko, ``{ABC:} an academic industrial-strength
  verification tool,'' in \emph{Computer Aided Verification}, 2010, pp. 24--40.

\bibitem{MP01}
A.~Mishchenko and M.~Perkowski, ``Fast heuristic minimization of exclusive
  sum-of-products,'' in \emph{Int'l Reed-Muller Workshop}, 2001.

\bibitem{AGM14}
L.~G. Amar{\`{u}}, P.~Gaillardon, and G.~De~Micheli, ``Majority-inverter graph:
  {A} novel data-structure and algorithms for efficient logic optimization,''
  in \emph{Proc. DAC}, 2014, pp. 194:1--194:6.

\bibitem{SFWD12}
M.~Soeken, S.~Frehse, R.~Wille, and R.~Drechsler, ``{RevKit}: {A} toolkit for
  reversible circuit design,'' \emph{Multiple-Valued Logic and Soft Computing},
  vol.~18, no.~1, pp. 55--65, 2012.

\bibitem{TVM16}
H.~Thapliyal, T.~Varun, and E.~Munoz-Coreas, ``{Quantum circuit design of
  integer division optimizing ancillary qubits and $T$-count},'' \emph{arXiv},
  vol. 1609.01241, 2016.

\bibitem{CDK+04}
S.~A. Cuccaro, T.~G. Draper, S.~A. Kutin, and D.~P. Moulton, ``A new quantum
  ripple-carry addition circuit,'' \emph{arXiv quant-ph/0410184}, 2004.

\bibitem{Maslov16}
D.~Maslov, ``Advantages of using relative-phase {Toffoli} gates with an
  application to multiple control {Toffoli} optimization,'' \emph{Phys. Rev.
  A}, vol.~93, p. 022311, 2016.

\bibitem{BBC+95}
A.~Barenco \emph{et~al.}, ``Elementary gates for quantum computation,''
  \emph{Phys. Rev. A}, vol.~52, pp. 3457--3467, 1995.

\end{thebibliography}

\end{document}